\documentclass[proof]{WileyASNA-v1}

\articletype{Article Type}%

\received{13 October 2020}
\revised{18 October 2020}
\accepted{22 October 2020}

\begin{document}

\title{A roadmap to strange star}
\author[aff4,aff5]{Renxin Xu*}
\author[aff1,aff2]{Xiaoyu Lai}
\author[aff3]{Chengjun Xia}

\address[aff4]{School of Physics and State Key Laboratory of Nuclear Physics and Technology, Peking University, Beijing 100871, China}
\address[aff5]{Kavli Institute for Astronomy and Astrophysics, Peking University, Beijing 100871, China}
\address[aff1]{Department of Physics and Astronomy, Hubei University of Education, Wuhan 430205, China}
\address[aff2]{Research Center for Astronomy, Hubei University of Education, Wuhan 430205, China}
\address[aff3]{School of Information Science and Engineering, Zhejiang University Ningbo Institute of Technology, Ningbo 315100, China}

\corres{*Renxin Xu, School of Physics, Peking University, Beijing 100871, China. \email{r.x.xu@pku.edu.cn}}

\abstract{%
What if normal baryonic matter is compressed so tightly that atomic nuclei come into close contact?
This question has been asked since 1930s. The fist answer was presented by Lev Landau whose speculation has been developed, and the concept of neutron star is then popularized.
However, another answer is related to strange star, which becomes worthy of attention especially after the establishment of the standard model of particle physics in 1960s.
The basic ideas of this study are introduced pedagogically. We must point out emphatically that flavour symmetry of and strong coupling between quarks would be essential in seeking true answer to the question.
The final answer is expected to appear in the era of multimessenger astronomy.
It is emphasized too that, besides the differences of global properties (e.g., mass-radius relation, maximum mass, tidal deformability), the strong-bound surface of strange star (rather than the gravity-bound one for conventional neutron star) could play an important role in identifying a strange star by astronomical observations.
}%

\maketitle

\section{Introduction}
Literally, ``strange'' star could not exist because it is supposed to be unusual and ugly as its name indicates, and because Nature may love something beautiful.
However, in this short note, we would like to convince you that strange stars are symmetrical, to be even more beautiful than so-called ``neutron'' stars!
Therefore, one may think that a pulsar should actually be a strange star if Nature really loves beauty.

\begin{figure*}[h!]
\centering
\includegraphics[scale=0.5]{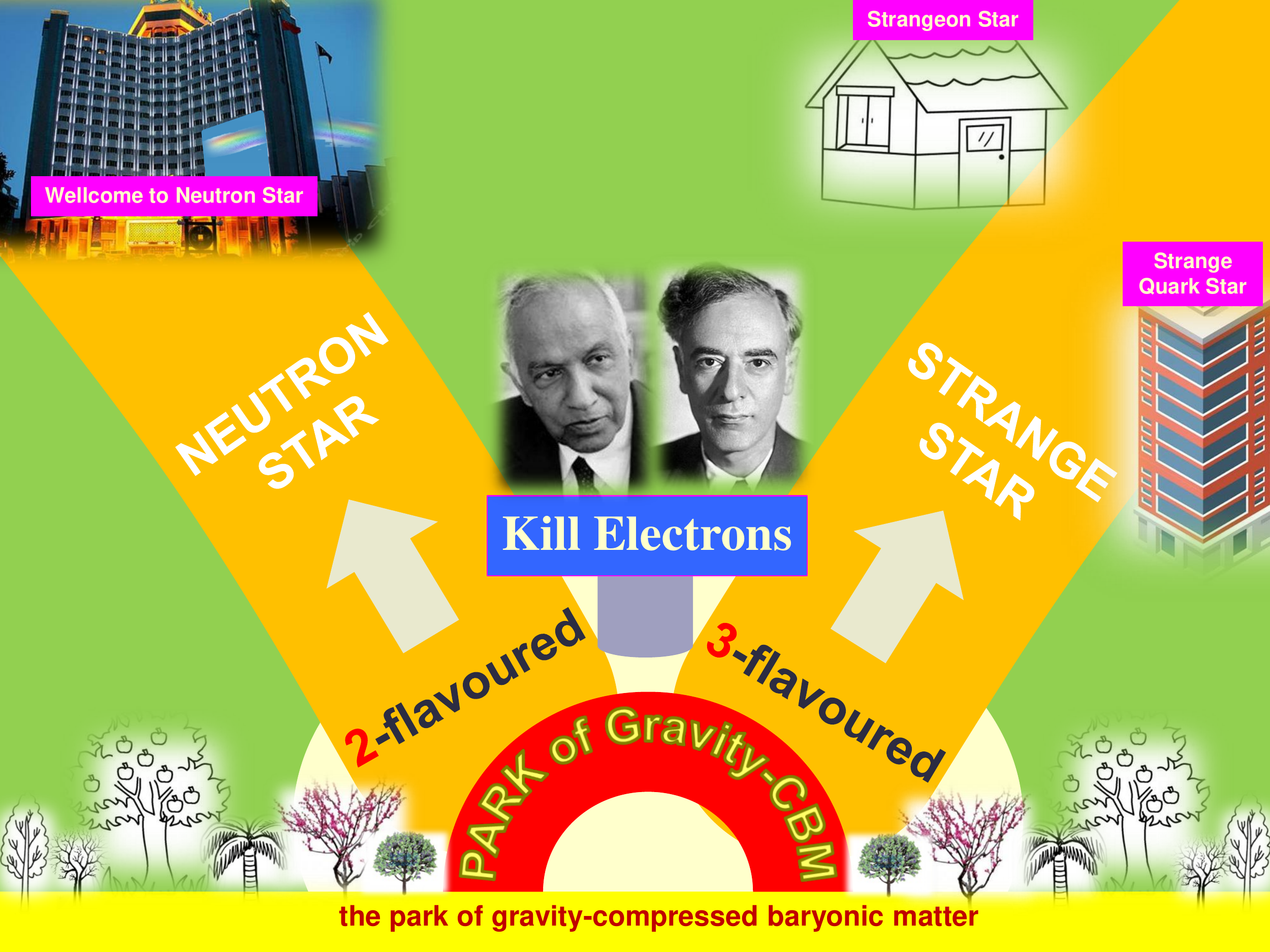}
\caption{A roadmap for the gravity-compressed baryonic matter (CBM) created after a core-collapsed supernova of an evolved massive star.
There are two sculptures in front of the gate that you just come in, but we have to choose one of the two ways to go: either the way to neutron star or the other to strange star. Most of the tourist go to the left side to see ``neutron star'', but the landscape could be more beautiful if you go to the right side.
} %
\label{fig:roadmap}
\end{figure*}

One of the puzzling problems, to be solved in today's multimessenger era of astronomy, could be related to the real nature of gravity-compressed baryonic matter (CBM) created after a core-collapsed supernova of an evolved massive star.
The remnant could be a neutron star (nucleon star, neutron-rich), a strange quark star (light-flavour quarks as the degree of freedom), or even a strangeon star (similar to nucleon star, but with light-flavour symmetry of quarks), all of which are explained approachably in this essay.
As tourist guiders, we would introduce you these objects in this CBM park (Fig.\ref{fig:roadmap}), and you would judge by yourself which ``building'' you prefer to live in.

\section{Historical notes: from Chandrasekhar to Landau}

Our story starts from the fact that more electrons e$^-$ (rather than positrons e$^+$) participant in the world because the lightest flavours of quarks (up and down), with equal numbers, are charged positively, and the lightest lepton charged, electron, has to come in for neutrality.
This is the reason that we have an atomic nucleus positively charged, while electrons outside are surrounding the nucleus due to the electromagnetic interaction relatively weaker than the strong one.
This form of normal atom matter is fine, but what if such baryonic matter is squeezed by gravity so greatly that atomic nuclei come into close contact? --- Aha, an interesting problem in astrophysics!

Rational thinking about gravity-compressed baryonic matter (CBM) dates back to the 1930s when Chandrasekhar and Landau were active in science.
In Fig.~\ref{fig:roadmap}, you will see two sculptures just inside the door from which you come in: Subrahmanyan Chandrasekhar (1910-1995) and Lev Landau (1908-1968). Their thoughts are relevant to the asymmetry of e$^\pm$, to be larger as density increases when normal baryonic matter is squeezed extremely by gravity in an evolved star.

Chandrasekhar proposed that the electron (our world is full with electron e$^-$, rather than positron e$^+$) degenerate pressure of a dead star (i.e., nuclear power ceases inside) would not be able to keep stand against its self-gravity if it's mass is higher than a critical value, now so-called the Chandrasekhar limit.
It is well know that radiative and thermal pressures balance the gravity in popular main-sequence stars in which nuclear fusion of light nuclei power stellar radiation. But, what if the nuclear energy source was exhausted?
As an underground student, major in physics, Chandrasekhar was interested in the Fermi-Dirac statistics, and, as still a teenager, published his first scientific paper~\citep{1929RSPSA.125..231C} on Compton scattering of moving electrons which obey the Fermi-Dirac statistics, to supplement Dirac's work on Compton scattering of moving electrons with a Maxwellian distribution in hot stellar atmosphere~\citep{1925MNRAS..85..825D}.
Certainly, Chandrasekhar was most intrigued by Fowler's work on the constitution of white dwarf stars~\citep{1926MNRAS..87..114F}, with ``...so that densities up to $10^{14}$ times that of terrestrial material may not be impossible'' (i.e., the nuclear density is possible) in the {\it Introductory}.
Upon graduation, Chandrasekhar was on a boat to UK for postgraduate study advised by Fowler at the University of Cambridge, where he tried to combine Fowler's work with Einstein's theory of special relativity.
His efforts returns success, with finding that this combination predicted a mass limit of white dwarfs, finally published in an American journal ``{\it the Astrophysical Journal}'' in 1931~\citep{1931ApJ....74...81C}.
However, what if a dead star has a mass higher than the Chandrasekhar limit? Landau presented the first answer to this.

Landau was thinking that an extremely large asymmetry of e$^\pm$ would result in a state of matter neutron-rich in our universe, exactly in the case of a dead star beyond the Chandrasekhar limit, though he thought that the conclusion would also be correct even for an active star (i.e., a star with nuclear power), writing ``{\it We expect that this must occur when the density of matter becomes so great that atomic nuclei come in close contact, forming one gigantic nucleus}'' in his paper~\citep{1932PhyZS...1..285L} before Chadwick's discovery of neutron~\citep{2013PhyU...56..289Y}.
From a view point of today's physics, neutronization occurs as e$^-$-density increases, with a final product of neutron-rich giant nucleus,\footnote{%
In an anti-cosmos, nucleus is negatively charged, and an extremely high density of positron would also result in a process of antineutronization: $e^++{\bar p}\rightarrow {\bar n}+{\bar \nu}_e$.
}%
\begin{equation}
    e^-+p\rightarrow n+\nu_e.
\label{neutronization}
\end{equation}
It is well known that Landau was one of the greatest physicists, especially in condensed matter physics (the theory of superfluidity), but Landau did care about his idea of gigantic nucleus.
According to ``{\it Complete list of L D Landau's works}'' provided by Aksenteva~\cite{Aksenteva1998}, Landau published totally six {\it Nature} papers but three were one-authored only by himself:

\begin{itemize}
	\item[(1)] L. Landau, ``Origin of stellar energy'', {\it Nature}, 141, 333 (1938)
    \item[(2)] L. Landau, ``The theory of phase transitions'', {\it Nature}, 138, 840 (1936)\\
    Brief message of ``ZETF 7 (1937) 19, 627; Phys. Z. Sowj. 11 (1937) 26, 545''
    \item[(3)] L. Landau, ``The intermediate state of supraconductors'', {\it Nature}, 141, 688 (1938)\\
    Brief message of ``ZETF 13 (1943) 377; J. Phys. USSR 7 (1943) 99''
\end{itemize}

\noindent{%
The first one was actually based upon the publication~\cite{1932PhyZS...1..285L} published in 1932, and the latter two were summaries of previous works that might lead to his Nobel prize in physics in 1962.
Why did Landau addressed again his idea about gigantic nucleus and stellar energy?
It was said that Landau was submitting the manuscript to {\it Nature} in order to stand against his political pressure in 1937.
The paper was published finally in 1938, but Landau was still jailed (he was in prison from 28 April 1938 to 29 April 1939).
One can then see Landau's interests of stars from this real story.}

\section{The way to conventional neutron star}

Landau showed us that a giant nucleus would be neutron-rich, but did not tell us how large a giant nucleus is.
We may answer this question: the critical length could be the Compton wavelength of electron, $\lambda_{\rm c}=h/(m_{\rm e}c)=2.4\times 10^3$ fm, since electron becomes relativistic if it is confined in a scale of the Compton wavelength.
The critical baryon number could then be $A_{\rm c}\sim \lambda_{\rm c}/$fm$^3\sim 10^9$ as the volume of a baryon is order of fm$^3$.

Bigger is different!
For an atomic nucleus with length scale of ($10^0-10^1$) femtometres, electrons have to be outside the nucleus because they don't feel the strong force of quarks and gluons, and should usually be non-relativistic.
But for a giant nucleus with length scale $\gtrsim \lambda_{\rm c}$, electrons should be inside and relativistic because of the electromagnetic interaction between electrons and quarks.
Let's see the huge energy of energetic electrons by a simple exercise of squeezing an apple, since the electron kinematic energy increases as the density becomes higher and higher.
The total baryon number of an apple is $A_{\rm apple}\sim 100$g$/u\sim 10^{26}\gg A_{\rm c}$, with $u$ the mass unit.
Electrons are non-relativistic in normal matter before squeezing, but they should be extremely relativistic since the giant nucleus of ``squeezed'' apple is $\sim 0.5 \mu$m $\gg \lambda_{\rm c}$, at nuclear density $\rho_{\rm n}=0.16$ fm$^{-3}$, with electron Fermi energy $E_{\rm e}\simeq (3\pi^2)^{1/3}\hbar c~(\rho_{\rm n}/2)^{1/3} \sim 300$ MeV if electrons keeps without conversion by the weak interaction (\ref{neutronization}).
Note that the mass difference of neutron and proton is only $\sim 1.3$ MeV $\ll E_{\rm e}$, and that the collapsed system at nuclear density would be unstable because of energetic electrons.
Landau's idea to cut down the system energy is to kill electrons via neutronization of the reaction~(\ref{neutronization}).
This is certainly also effective for a gravity-squeezed core inside an evolved massive star, but with baryon number $A_{\rm star}\simeq M_\odot/u\sim 10^{57} \gg A_{\rm apple}$ so that gravity is not able to be negligible.
Because neutrons decay into protons in vacuum with zero electron-density, a neutron star has to be covered by a crust, at the bottom of which the electron-density is high enough to prohibit the decay.
This implies that free neutrons cannot exist on the surface of a neutron star, with an extremely low density compared to the nuclear density, and that the mass of conventional neutron star should be $>0.1M_\odot$.

However, two uncertainties exist in the study of neutron stars: Quarks confined or deconfined? Strangeness significant or not?
These are the topics of next section.

\section{The way to strange star}

Let's consider {\it if strangeness is significant} at first.
The typical energy scale of dense matter around nuclear density is much larger than the masses of light-flavour quarks but is smaller than that of heavy-flavour ones, and another possibility of killing electrons for giant nucleus is thus provided if 3-flavour (u, d and s quarks) symmetry is restored.
We can see the energy scale and its impacts as following.
For strong matter at a few nuclear densities, the separation between quarks is $\Delta x\sim 0.5$ fm, and the energy scale is thus of order $E_{\rm scale}\simeq \hbar/\Delta x\sim 0.5$ GeV, according to Heisenberg's relation\footnote{%
Note that the $E_{\rm scale}$-energy depends seemingly on quark-number density of strong condensed mater at zero pressure, but on the coupling strength of fundamental strong interaction in fact~\citep{xu18}.
}. %
Note that the mass difference between strange and up/down quarks is only $\Delta m_{\rm uds}\sim 0.1$ GeV.
We can then know that strangeness should be included to reveal the secret of giant nucleus, even normal atomic one, which has already been noted since 1970s.
However, it has always been wondered why the stable nuclei in the world are 2-flavored. We may provide a simple answer: normal atomic nuclei are too small to have a 3-flavor symmetry, but this does not apply to a giant nucleus!
The Fermi energy of electrons is negligible for micro-nuclei but is significant for a gigantic-nucleus produced in the core of a massive star during supernova.
Conventionally, neutronization has been the explanation for the removal of energetic electrons even since Landau, but an alternative explanation could be strangenization, i.e., restoration of 3-flavour symmetry with approximately equal number densities of u, d and s quarks.

Secondly, we consider the other question: {\it could quarks be deconfined in compact stars}?
The perturbative QCD, based on asymptotic freedom, works well at energy scale of $\Lambda_\chi>1$ GeV, therefore, one has $\Delta m_{\rm uds}\ll E_{\rm scale}<\Lambda_\chi$.
This fact may have impacts on the nature of strong matter.
(1). Chiral symmetry could be broken and quarks would be dressed with mass $\tilde{m}_{\rm q}\sim 0.3$ GeV, as is evident from both approaches of lattice-QCD and of Dyson-Schwinger equations.
(2). The coupling could still be strong, even with constant $\alpha_{\rm s}\gtrsim 1$.
It was then suggested that quark clustering occurs in realistic cold dense matter around the nuclear density because of strong coupling between quarks and gluons~\citep{Xu2003,Xu2009,LX2009}.
The quark cluster is actually nucleon (i.e., proton and neutron) in case of two flavours of quarks (up and down), but is renamed ``{\it strangeon}'' for the strong matter with three flavour symmetry (up, down and strange).
A strangeon star could be in a solid state when its temperature is much lower the the interaction energy between strangeons, this model could help us understand different manifestations observed in pulsar-like stars~\citep{LX2017}.

Although nucleons are non-strange, in modern physics, it is worth emphasizing that, because of the asymmetry of
$e^-$ and $e^+$, virtual {\it strange} quarks in the nucleon
sea could materialize as valence ones when normal baryonic
matter in the core of an evolved massive star is squeezed so great
that nuclei come in close contact.
For lepton-related weak interactions of $u+e^-\rightarrow s+\nu_e$ and ${\bar u}+e^+\rightarrow {\bar s}+{\bar \nu}_e$, the former should be more effective than the latter in dense matter at higher and higher density, producing eventually valence strange quarks as many as the up and down quarks.
An alternative way is for two flavours only ($u+e^-\rightarrow d+\nu_e$ and ${\bar u}+e^+\rightarrow {\bar d}+{\bar \nu}_e$), resulting in an extremely asymmetric state of isospin.
In a word, in a 2-flavoured way, with the freedom degree of nucleon, CBM should be {\it asymmetric} neutron-rich to the building of neutron star; however, in a 3-flavoured way, we have {\it symmetric} either strange quark star~\citep{Witten1984,Haensel1986,Alcock1986} or strangeon star~\citep{Xu2003}, as illustrated in Fig.~\ref{fig:roadmap}.

\begin{figure}[h!]
\centering
\includegraphics[scale=0.5]{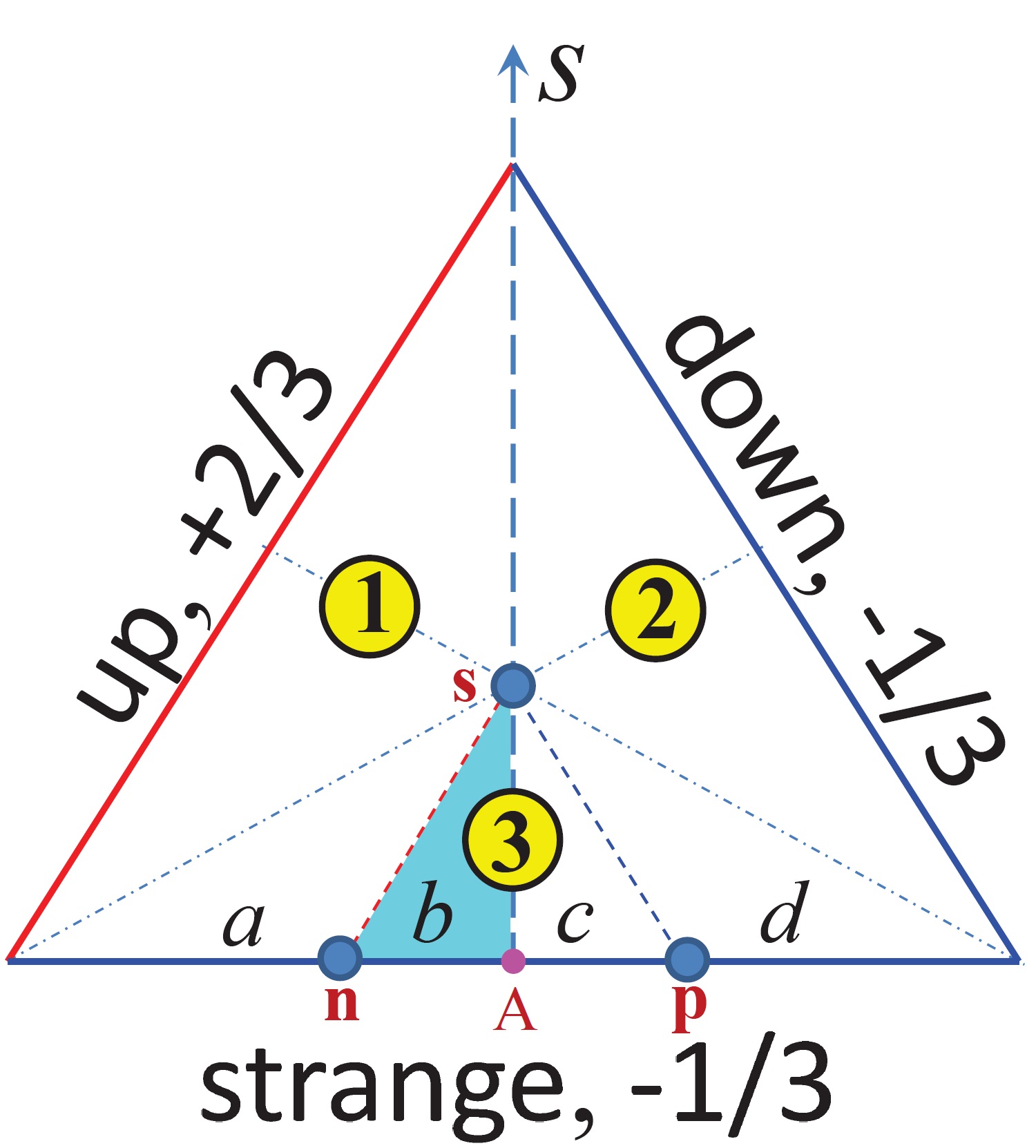}
\caption{Triangle of light-quark flavours. The point inside the triangle defines a state with certain quark numbers of three flavours (\{$n_{\rm u}, n_{\rm d}, n_{\rm s}$\} for up, down and strange quarks), which are measured by the heights of the point to one of the triangle edges.
Point ``s'' is the center of the triangle, at which one has $n_{\rm u}=n_{\rm d}=n_{\rm s}$.
Line ``sn'' is parallel to the up edge, while line ``sp'' to the down edge.
Axis $S$ is for strangeness, where the isospin symmetry is also perfect.
} %
\label{fig:triangle}
\end{figure}

In the light that nature might love symmetry, one may take an advantage of a triangle diagram as in Fig.~\ref{fig:triangle}.
Due to baryon conservation, it is convenient to discuss the quark numbers of the three flavours there, for a given baryon density, $n_{\rm b}=(n_{\rm u}+n_{\rm d}+n_{\rm s})/3$, with quark number density of up $n_{\rm u}$, down $n_{\rm d}$ and strange $n_{\rm s}$.
It is evident that the bottom strange edge is divided into three equal parts
by points ``n'' and ``p''  because the triangle ``$\triangle$snp''
is left-right symmetrical to the ``$S$''-axis but shrinks by
two-thirds.
Normal nuclei are around point ``A'', conventional neutron stars in point ``n'' while extremely unstable proton stars in point ``p'', but strange stars (both quark star and strangeon star) in point ``s''.

We emphasize that the flavour-asymmetric point ``n'' should led to the existence of normal atomic matter on the gravity-bound surface of convectional neutron star, while the flavour-symmetric point ``s'' would result in a ``bare'' surface (ie., strong-bound surface, but possibly being covered by a crust if significant accretion process occurs).
For point ``n'', due to the large asymmetry of neutron and proton (ie., the isospin asymmetry, essentially the asymmetry of up and down quarks), a high number density of electrons would be necessary to suppress the $\beta$-decay of neutron to proton. Normal atom matter bound by gravity on surface could meet the standard of such an electron density, and the conventional neutron stars are therefore gravity-bound on surface, which usually have smaller radii with larger masses~\citep{wu+2020}.
For point ``s'', in an analogy to stable atomic nuclei with two-flavour symmetry, three-flavoured strange matter is supposed to be absolutely stable on surface, i.e. self-confined by strong force.
The sharp difference, either atom matter or quark/strangeon matter on surface, would make dissimilarity of pulsar magnetospheric activities, consequently~\citep{xu+1999,Lu+2019}.
Additionally, stellar radius becomes usually larger as the mass increases for such strong-bound strange quark star and strangeon star.

To examine strangeon matter with a more detailed microscopic dynamics, recently we have developed a linked bag model~\citep{Miao2020}, where the strong interaction is treated effectively via quark propagation between separated bags. With the model parameters carefully adjusted to reproduce the saturation properties of nuclear matter, the possible existence of strangeon matter and strangeon star were analyzed. It was shown that the maximum mass of strangeon stars can be as large as $\sim 2.5M_\odot$, while the tidal deformability of a $1.4M_\odot$ strangeon star lies in the range of $180\lesssim \Lambda_{1.4} \lesssim 340$, which is in consistent with pulsar observations. Certainly, the maximum mass could be $\gtrsim 2.5M_\odot$ if the parametric $N_{\rm q}$- and $f$-values increase. More micro-physical efforts in modelling strangeon matter is surely welcome.

Besides the theoretical study, new achievements of strangeon star study are of astrophysical implications, which could provide observational evidence for strangeon stars, including the positive $P_2-P_2$ correlation of PSR B2016+28 with drifting subpulses~\citep{Lu+2019} and the magnetospheric origin of fast radio burst~\citep{Wang+2018,Jiang+2020,2019ApJ...876L..15W,2020ApJ...899..109W,luo+20}, and even the glitch behavior~\citep{Zhou+2014,Lai+2018,Wang20+}.
The biggest single-dish radio telescope in the world, i.e., the China's Five-hundred-meter Aperture Spherical radio Telescope (FAST), is going to regularly observe pulsar-like compact stars, with extremely high sensitivity but without requiring the complicated data processing required for an antenna array.
We may then anticipate a FAST~\citep{Jiang19+} era of pulsar-related science to come.

\section{Summary}

In this note, we try to show you the essence of different ideas about the nature of compressed baryonic matter produced in gravity-squeezed core of massive star, and try to convince you that the basic units inside pulsar-like compact stars could be 3-flavour symmetric strangeons, rather than 2-flavour asymmetric nucleons, if Nature really loves symmetry when building the world.
Certainly, we are expecting to test the strangeon star model further in the future, especially taking advantage of the FAST.

\section*{ACKNOWLEDGMENTS}
The author would like to thank those involved in the continuous discussions in the pulsar group at Peking University. This work is supported by the National Key R\&D Program of China (No. 2017YFA0402602), the National Natural Science Foundation of China (Grant Nos. 11673002, U1831104), and the Strategic Priority Research Program of CAS (No. XDB23010200). XL is also supported by the Science and technology research project of Hubei Provincial Department of Education (No. D20183002).

\bibliography{references}

\end{document}